\documentclass[letterpaper,preprintnumbers,prd,twocolumn,nofootinbib,nobibnotes,showpacs]{revtex4}
\usepackage{amsfonts}
\usepackage{mathrsfs}
\usepackage{epsfig}
\usepackage{graphicx}%
\usepackage{dcolumn}
\usepackage{amsmath}
\usepackage{color}
\usepackage{color}
\makeatletter
\def\btt#1{\texttt{\@backslashchar#1}}%
\DeclareRobustCommand\bblash{\btt{\@backslashchar}}%
\makeatother
\begin{document}

\title{Exact solutions in $F(R)$ theory of gravity}
\author{Changjun Gao}\email{gaocj@bao.ac.cn} \affiliation{Key Laboratory for Computational Astrophysics, National Astronomical
Observatories, Chinese Academy of Sciences, Beijing, 100012,
China} \affiliation{State Key Laboratory of Theoretical Physics,
Institute of Theoretical Physics, Chinese Academy of Sciences,
Beijing 100190, China}
\author{You-Gen Shen}
\email{ygshen@center.shao.ac.cn} \affiliation{Shanghai
Astronomical Observatory, Chinese Academy of Sciences, Shanghai
200030, China}

\date{\today}

\begin{abstract}
We find a new method for looking for the static and spherically symmetric solutions in $F(R)$ theory of gravity.
With this method, a number of new solutions in terms of the analytic functions are obtained. We hope this investigation may be of some help in
the searching for some other solutions in $F(R)$ theory of gravity.

 \end{abstract}


\maketitle

\section{Introduction}
The searching for exact solutions in the $F(R)$ theory of gravity is important but challenging because the equations of motion are
of fourth order derivatives. Although the searching is challenging, many remarkable achievements have been made. We here briefly review these achievements.

For the specific choice of $R^{1+\delta}$, a class of exact static spherically symmetric solution has been presented in \cite{john:05}.
Using the method of Lagrange multiplier, reference \cite{seb:11} presents a Lagrangian derivation of the equation of motion for the static spherically symmetric spacetimes in $F(R)$ theory of gravity. It is found the corresponding equations of motion are simply of first order derivative and thus some new solutions are obtained.
Reference \cite{seb:12} constructs some new static spherically symmetric interior solutions in $R^{1+\delta}$ theory.
With the method of Noether symmetries,  reference \cite{seb:13} find some new static spherically symmetric solutions in $F(R)$ theory.
Reference \cite{seb:14} constructs the spherically symmetric solutions of $F(R)$ gravity using their input function method.
Reference \cite{seb:15} presents a static axially symmetric vacuum solution for $F(R)$ gravity in the Weyl's canonical coordinates. Finally, using the so-called generator method, reference \cite{ami:15} presents some static spherically symmetric solutions in $F(R)$ theory in $n$ dimensional spacetimes.  For more works on static solutions in $F(R)$ gravity, we refer to the references [7-29].

In this paper, we develop an alternative method for looking for static and spherically symmetric solutions in $F(R)$ theory of gravity. Similar to reference \cite{seb:14}, we appropriately fix one of the unknown functions initially. Then the equations of motion become remarkably simple and solvable.
Using this method, a number of new solutions in terms of known analytic functions are obtained. We hope the investigation may be of some help in
the searching for other new solutions in $F(R)$ theory of gravity with or without energy-momentum tensor. Throughout this paper, we adopt the
system of units in which $G=c=\hbar=1$ and the metric signature
$(-,\ +,\ +,\ +)$.
\section{exact solutions}\label{sec:split}
\subsection{Ansate for the metric}\label{sec:split}
In general, the metric of a four dimensional, static and spherically symmetric spacetime takes the form of
\begin{equation}
   ds^2=-U\left(x\right)dt^2+\frac{1}{A\left(x\right)}dx^2+f\left(x\right)d\Omega^2\;,
\end{equation}
where $x$ plays the role of radial variable and $d\Omega^2$ is the line element for the two dimensional unit sphere. $U(x),\ A(x),\ f(x)$ are three functions to be determined.  One can always express $A(x)$ as follows
\begin{equation}
  A\left(x\right)=U\left(x\right)B\left(x\right)f\left(x\right)^2\;,
\end{equation}
 with $B\left(x\right)$ a new function to be determined. The metric becomes
 \begin{equation}
   ds^2=-U\left(x\right)dt^2+\frac{1}{U\left(x\right)B\left(x\right)f\left(x\right)^2}dx^2+f\left(x\right)d\Omega^2\;.
\end{equation}
 Let
 \begin{equation}
  r=\int\frac{dx}{\sqrt{B}}\;,
\end{equation}
 then we find the metric can be written as
 \begin{equation}
   ds^2=-U\left(r\right)dt^2+\frac{1}{U\left(r\right)f\left(r\right)^2}dr^2+f\left(r\right)d\Omega^2\;.
\end{equation}
 The advantage of this form is that the determinant of the metric is independent of the radial variable, $r$. We shall see later the equations of motion are greatly simplified with this form of metric.
Given the metric, the Ricci scalar $R$ is calculated to be
\begin{equation}
R=\frac{3}{2}Uf^{'2}+3fU^{'}f^{'}+2fUf^{''}+f^2U^{''}-\frac{2}{f}\;,
\end{equation}
where the prime denotes the derivative with respect to $r$.

\subsection{Exact solutions}\label{sec:split}

The Lagrangian density of $F(R)$ theory of gravity takes the form
\begin{equation}
\mathscr{L}=F\left(R\right)\;.
\end{equation}

Compared to struggling with the Einstein equations, it is much more convenient to directly vary the action
\begin{equation}
S=\int d^4 x\sqrt{-g}\mathscr{L}\;,
\end{equation}
with respect to $U$ and $f$, respectively. By so doing, we find the equations of motion are
\begin{eqnarray}\label{eq:hessence}
&&\left(f^2K\right)^{''}-\left(3fKf^{'}\right)^{'}+K\left(\frac{3}{2}f^{'2}+2ff^{''}\right)=0\;,\\
&&\left(2fUK\right)^{''}-\left[K\left(3Uf\right)^{'}\right]^{'}\nonumber\\&&+K\left(3U^{'}f^{'}+2Uf^{''}+2fU^{''}+\frac{2}{f^2}\right)=0\;,\\
&&K=K\left(r\right)\equiv F_{,R}\;.
\end{eqnarray}
Here $F_{,R}\equiv\frac{dF}{dR}$. We have only two equations of motion (9) and (10) but three functions to be determined: $U$, $f$ and $K$. Therefore the system of equations (9) and (10)
are not closed. It seems reasonably to initially specify a certain form of $F(R)$ by hand. However, by this way, the equations of motion
turn out to be fourth order differential equations and it is very difficult to find their solution.

Observing equations (9) and (10), we see if we fix not the expression of $F(R)$, but the expression of $K(r)$ initially, then we can solve for $f(r)$ from Eq.~(9). Once $K(r)$ and $f(r)$ are given, the function $U(r)$ can be obtained from Eq.~(10). Above is exactly the strategy what we shall take in the following calculations.

In order to obtain the analytic solutions, we specify $K$ as follows
\begin{equation}\label{scaling1}
\begin{split}
    &K\left(r\right)=f\left(r\right)^{\alpha}\;.
\end{split}
\end{equation}
Then we obtain a class of exact solutions.
\begin{equation}\label{scaling1}
\begin{split}
    &f\left(r\right)=b_0r^{\frac{2\left(\alpha+1\right)}{2\alpha^2+2\alpha+3}}\;,\\
    &K\left(r\right)=b_0^\alpha r^{\frac{2\alpha\left(\alpha+1\right)}{2\alpha^2+2\alpha+3}}\;,\\
    &U\left(r\right)=d_0r^{\frac{2\alpha\left(2\alpha-1\right)}{2\alpha^2+2\alpha+3}}+c_1r^{-\frac{4\alpha+1}{2\alpha^2+2\alpha+3}}+c_2r^{\frac{2\alpha+2}{2\alpha^2+2\alpha+3}}\;,\\
    &R=d_1r^{-\frac{2\alpha+2}{2\alpha^2+2\alpha+3}}+d_2r^{-\frac{2\alpha\left(2\alpha-1\right)}{2\alpha^2+2\alpha+3}}\;,\\
    &d_0\equiv-\frac{\left(2\alpha^2+2\alpha+3\right)^2}{\left(2\alpha^2-2\alpha-1\right)\left(4\alpha^2+2\alpha+1\right)b_0^3}\;,\\
    &d_1\equiv-6\frac{\alpha\left(\alpha-1\right)}{b_0\left(2\alpha^2-2\alpha-1\right)
    }\;,\\
    &d_2\equiv -6\frac{c_2b_0^2\left(2\alpha^3-\alpha^2-5\alpha-2\right)}{\left(2\alpha^2+2\alpha+3\right)^2}\;,\\
\end{split}
\end{equation}
where $\alpha, b_0, c_1, c_2$ are constants. We see the expression of $K(r)$ and $R(r)$ are obtained. Taking Eq.~(11) into consideration,
we can derive the expression of $F(R)$. In the next, we shall investigate some specific solutions with different exponent $\alpha$ in order that the expression of $F(R)$ are analytic functions.

To understand the solutions more explicitly, let's rewrite the metric in the Schwarzschild coordinate system. Rescale proper distance $s$ and time variable $t$ as follows, $s^2\longrightarrow b_0s^2$,$t^2\longrightarrow b_0t^2$ and let
\begin{eqnarray}
r&=&x^{\frac{2\alpha^2+2\alpha+3}{\alpha+1}}\;,
\end{eqnarray}
we obtain
\begin{eqnarray}
&&ds^2=-x^{\frac{4\alpha^2-2\alpha}{\alpha+1}}\left[d_0
    +c_1x^{-\frac{4\alpha^2+2\alpha+1}{\alpha+1}}+c_2x^{-\frac{4\alpha^2-4\alpha-2}{\alpha+1}}\right]dt^2\nonumber\\&&
    +d_3\left[d_0
    +c_1x^{-\frac{4\alpha^2+2\alpha+1}{\alpha+1}}+c_2x^{-\frac{4\alpha^2-4\alpha-2}{\alpha+1}}\right]^{-1}dx^2\nonumber\\&&+x^2d\Omega^2\;,
\end{eqnarray}
with
\begin{eqnarray}
&&d_3\equiv\frac{1}{b_0^3}\frac{\left(2\alpha^2+2\alpha+3\right)^2}{\left(\alpha+1\right)^2}\;,\\
&&K\left(x\right)=x^{2\alpha}\;,\\
&&R=b_0d_1x^{-2}+b_0d_2x^{\frac{2\alpha\left(1-2\alpha\right)}{\alpha+1}}\;.
\end{eqnarray}
Rescale $t,\ c_1,\ c_2$ as follows
\begin{eqnarray}
t^2\rightarrow t^2/d_3\;,\ \ c_1\rightarrow c_1d_3\;,\ \ c_2\rightarrow c_2d_3\;,
\end{eqnarray}
the metric is simplified to be
\begin{eqnarray}
&&ds^2=-x^{\frac{4\alpha^2-2\alpha}{\alpha+1}}\left[c_0
    +c_1x^{-\frac{4\alpha^2+2\alpha+1}{\alpha+1}}+c_2x^{-\frac{4\alpha^2-4\alpha-2}{\alpha+1}}\right]dt^2\nonumber\\&&
    +\left[c_0
    +c_1x^{-\frac{4\alpha^2+2\alpha+1}{\alpha+1}}+c_2x^{-\frac{4\alpha^2-4\alpha-2}{\alpha+1}}\right]^{-1}dx^2\nonumber\\&&+x^2d\Omega^2\;,
\end{eqnarray}
where
\begin{eqnarray}
c_0\equiv\frac{d_0}{d_3}=-\frac{\left(\alpha+1\right)^2}{\left(2\alpha^2-2\alpha-1\right)\left(4\alpha^2+2\alpha+1\right)}\;.
\end{eqnarray}
We see there are three parameters in the solution, i.e., $\alpha,\ c_1,\ c_2$. We shall see later $c_1$ and $c_2$ play the role of mass-like term and $\lambda$(cosmological constant)-like term, respectively. Up to this point, we are able to understand the solutions and construct the expression of $F(R)$.
\subsection{some examples}
\subsubsection{$F=R-2\lambda$}
Consider the first term in the expression of $g_{00}$ and let it be a constant, we obtain $\alpha=0,\ \frac{1}{2}$.
In the case of $\alpha=0$, the solution is given by
\begin{eqnarray}\label{sch-de0}
&&ds^2=-\left[1
    +c_1x^{-1}+c_2x^2\right]dt^2+x^2d\Omega^2\nonumber\\&&+\left[1
    +c_1x^{-1}+c_2x^{2}\right]^{-1}dx^2\;,\nonumber\\
&&K\left(x\right)=1\;,\nonumber\\
&&F\left(R\right)=R-6c_2\;.
\end{eqnarray}
The metric is exactly the well-known Schwarzschild-de Sitter solution provided that
\begin{eqnarray}
c_2=\lambda/3\;, \ \ c_1=-2M\;.
\end{eqnarray}
Here $\lambda, M$ are the cosmological constant and the black hole mass respectively.
\subsubsection{$F=-2\sqrt{12c_2-R}$}
 When $\alpha=\frac{1}{2}$, we find the solution is given by
\begin{eqnarray}\label{12}
&&ds^2=-\left[{1}
    +c_1x^{-2}+c_2x^2\right]dt^2+\frac{x^2}{2}d\Omega^2\nonumber\\&&+\left[{1}
    +c_1x^{-2}+c_2x^{2}\right]^{-1}dx^2\;,\nonumber\\
&&K\left(x\right)=x\;,\nonumber\\
&&R={12}c_2-\frac{2}{x^2}\;,\nonumber\\
&&F\left(R\right)=-2\sqrt{12c_2-R}\;.
\end{eqnarray}
This is the massless Reissner-Nordstrom-de Sitter black hole with a deficit angle.
\subsubsection{$F=-\frac{\sqrt{2R}}{25}\left[9c_2-\sqrt{81c_2^2+20R}\right]
 \frac{\left[\sqrt{81c_2^2+20R}+6c_2\right]}{\left[\sqrt{81c_2^2+20R}+9c_2\right]^{\frac{1}{2}}}$}
 Let the second term in the expression of $g_{00}$ be a constant, we obtain $\alpha=-\frac{1}{4}$. In this case, the solution is given by
\begin{eqnarray}\label{sch-de}
 s^2&=&-x\left[{1}
    +\frac{c_1}{x}+c_2x\right]dt^2+2x^2d\Omega^2\nonumber\\&&+\left[{1}
    +\frac{c_1}{x}+c_2x\right]^{-1}dx^2\;,
 \end{eqnarray}
 \begin{eqnarray}\label{sch-de}
 &&F\left(R\right)=-\frac{\sqrt{2R}}{25}\left[9c_2-\sqrt{81c_2^2+20R}\right]
 \nonumber\\&&\left[\sqrt{81c_2^2+20R}+6c_2\right]\left[\sqrt{81c_2^2+20R}+9c_2\right]^{-\frac{1}{2}}\;,
\end{eqnarray}
which is also endowed with a deficit angle. When $c_2=0$, we have
\begin{eqnarray}
ds^2=-x\left({1}
    +\frac{c_1}{x}\right)dt^2+\left({1}
    +\frac{c_1}{x}\right)^{-1}dx^2+2x^2d\Omega^2\;.
\end{eqnarray}
Let $c_1=-2M$ with $M$ a positive constant. We have
\begin{eqnarray}
ds^2=-x\left({1}
    -\frac{2M}{x}\right)dt^2+\left({1}
    -\frac{2M}{x}\right)^{-1}dx^2+2x^2d\Omega^2\;.
\end{eqnarray}
This spacetime has an event horizon at $x=2M$ and a singularity at $x=0$ which is the same as the Schwarzschild spacetime.
But different from the schwarzschild spacetime, it has a deficit angle. Furthermore, the physical meaning of $M$
may be not the mass of some object and it is still unclear for us.
\subsubsection{$F=-\frac{81c_2^2}{R}-\frac{21}{2}$}
 Let the first term in the expression of Ricci scalar $R$ disappear, we obtain $\alpha=1$. In this case, the solution is given by
 \begin{eqnarray}\label{sch-de}
 ds^2&=&-x\left[1+\frac{c_1}{x^{\frac{7}{2}}}+c_2x\right]dt^2+\frac{4}{7}x^2d\Omega^2\nonumber\\&&+
 \left[1+\frac{c_1}{x^{\frac{7}{2}}}+c_2x\right]^{-1}dx^2\;,
\end{eqnarray}
\begin{eqnarray}\label{sch-de}
F\left(R\right)=-\frac{81c_2^2}{R}-\frac{21}{2}\;.
\end{eqnarray}
In this case, the spacetime has the same property as the case of $\alpha=-1/4$.
\subsubsection{$F=\frac{2}{9}\left(-R\right)^{\frac{3}{2}}$}
 When $\alpha=-\frac{1}{2}$, the solution is given by
\begin{eqnarray}
ds^2&=&-x^{4}\left[-{1}
    +\frac{c_1}{x^2}\right]dt^2+\frac{1}{2}x^2d\Omega^2\nonumber\\&&
    +\left[-{1}
    +\frac{c_1}{x^2}\right]^{-1}dx^2\;,
\end{eqnarray}
\begin{eqnarray}\label{sch-de}
F\left(R\right)=\frac{2}{9}\left(-R\right)^{\frac{3}{2}}\;.
\end{eqnarray}
This spacetime has an event horizon $x=\sqrt{c_1}$, a curvature singularity at $x=0$ and a deficit angle.
\subsubsection{$F=-\frac{16}{R}-\frac{30}{49}c_2$}
When $\alpha=2$, the solution is given by
\begin{eqnarray}
ds^2&=&-x^{4}\left[-1
    +\frac{c_1}{x^7}+\frac{c_2}{x^2}\right]dt^2+\frac{1}{7}x^2d\Omega^2\nonumber\\&&
    +\left[-1
    +\frac{c_1}{x^7}+\frac{c_2}{x^2}\right]^{-1}dx^2\;,
\end{eqnarray}
\begin{eqnarray}\label{sch-de}
F\left(R\right)=-\frac{16}{R}-\frac{30}{49}c_2\;.
\end{eqnarray}
The structure of this spacetime is the same as the case of $\alpha=-\frac{1}{2}$.
\subsection{general properties: singularity, event horizons and deficit angle}
In this section, let us study the structure of spacetime generally from Eq.~(15). We find in general, there are two event horizons and one curvature singularity in the spacetimes. Furthermore, a deficit angle always exists in the solutions except for the Schwarzschild-de Sitter solution. Now let us show these points one by one.

In the first place, equation (18) shows $x=0$ is always the singularity of spacetime because of the divergence of the first term  at $x=0$. Secondly, the spacetime has a deficit angle of $1/d_3$ except for the schwarzschild-de Sitter solution ($d_3$=1). In the following we turn to the horizons. The equation of event horizons could be derived from equation (15) as follows
\begin{eqnarray}
d_0+c_1x^{-\frac{4\alpha^2+2\alpha+1}{\alpha+1}}+c_2x^{-\frac{4\alpha^2-4\alpha-2}{\alpha+1}}=0\;.
\end{eqnarray}
Let $c_1, c_2$ absorb non-vanishing $d_0$ and
\begin{eqnarray}
k\equiv-\frac{4\alpha^2+2\alpha+1}{\alpha+1}, \ \ \  m\equiv-\frac{4\alpha^2-4\alpha-2}{\alpha+1},
\end{eqnarray}
we have
\begin{eqnarray}
1+c_1x^{k}+c_2x^{m}=0\;.
\end{eqnarray}
When
\begin{eqnarray}
\alpha<-1\;,
\end{eqnarray}
we find
\begin{eqnarray}
k>0\;, \ \ \ m>0\;.
\end{eqnarray}
On the other hand, when
\begin{eqnarray}
\alpha>-1\;,
\end{eqnarray}
$k, m$ would evolve with $\alpha$ as shown in Fig.~${1}$.
The structure of event horizons for $\alpha<-1$ is similar to the case of $\alpha>-1$. So
we may pay our attention to $\alpha>-1$.

In Fig.~1, we plot the evolution of exponents $k, m$ with respect to the dimensionless constants $\alpha$. When $\alpha=0$,
we have $k=-1$ and $m=2$ which corresponds to the Schwarzschild-de Sitter solution.

In Fig.~2, we plot the position of event horizons with respect to $\alpha$ for different $c_1$ and $c_2$. Curves $A,\ B,\ C,\ D, \ E$ correspond to $c_1=1.795,\ 1.995,\
2.039,\ 2.075,\ 2.175$ and $c_2=0.026, \ 0.031,\ 0.036,\ 0.039, \ 0.049$, respectively. The figure shows that, in the case of $A$, there exists two horizons for every $\alpha$. The smaller one represents the black hole event horizon and the larger one represents the cosmic event horizon. The case of $B$ is exactly the same as $A$. With the increasing of $c_1$ and $c_2$, the black hole event horizon expands and the cosmic event horizon shrinks. For some special value of $\alpha$, two horizons coincide which is the case of $C$. Continue increasing $c_1$ and $c_2$, we find both the black hole event horizon and cosmic event horizon vanishes and the black hole singularity is naked in some intervals of $\alpha$ (the case of $D$ and $E$). In this case, the conjecture of cosmic censorship is violated.
\begin{figure}[h]
\begin{center}
\includegraphics[width=9cm]{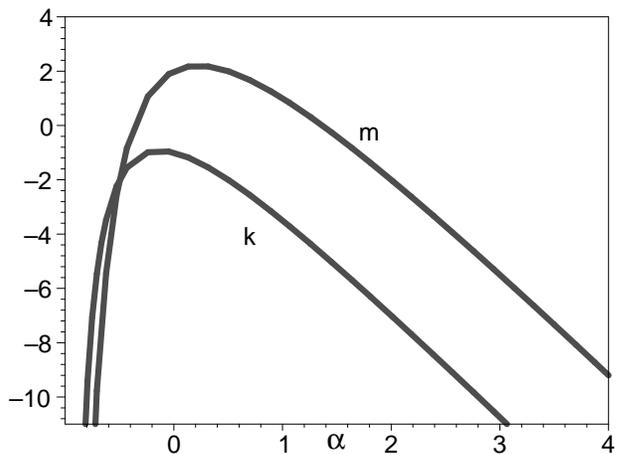}
\caption{The evolution of exponents $k, m$ with respect to the dimensionless constants $\alpha$. When $\alpha=0$,
we have $k=-1$ and $m=2$ which corresponds to the Schwarzschild-de Sitter solution. }. \label{e3}
\end{center}
\end{figure}

\begin{figure}[h]
\begin{center}
\includegraphics[width=9cm]{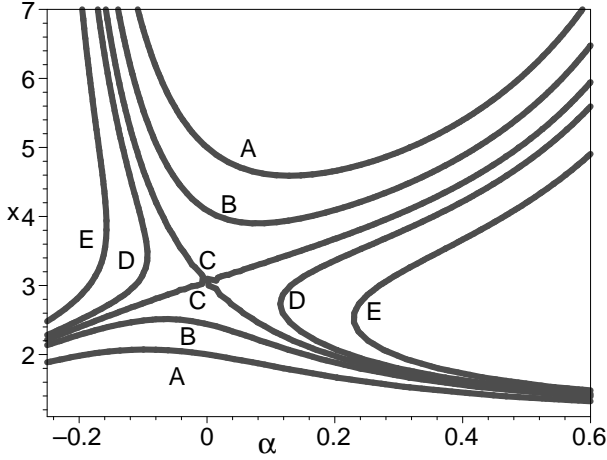}
\caption{The position of event horizons with respect to $\alpha$ for different $c_1$ and $c_2$. }. \label{e3}
\end{center}
\end{figure}
\section{Checking the solutions in the Jordan frame }
In section IIB, by varying the action with respect to $U$ and $f$, we derive two equations of motion, Eq.~(9) and
Eq.~(10). Then a class of solutions are obtained from the system of equations. Here we should emphasize that, in the applying of this method, some unnecessary solutions which actually
do not satisfy the Einstein equations might be present. The reason is simply. We obtain two equations of motion, Eq.~(9) and Eq.(10) with the variation method.
But Einstein equations show us with three equations of motion. Eq.~(9) and Eq.~(10) are exactly some combinations of the three Einstein equations. So solutions satisfied by the Einstein equations are certainly satisfied by Eq.~(9) and Eq.~(10). However, the converse may be not true. Therefore, one should check wether the solutions meet all the Einstein equations. In this section, we shall complete this examination.

In the Jordan frame, the generalized Einstein equations corresponding to action (7) take the form
\begin{eqnarray}
F_{,R}R_{\mu\nu}-\frac{1}{2}F g_{\mu\nu}-\left(\nabla_{\mu}\nabla_{\nu}-g_{\mu\nu}\nabla^2\right)F_{,R}=0\;.
\end{eqnarray}
In the Schwarzschild coordinate system, the metric for the static and spherically symmetric spacetime is
\begin{eqnarray}
ds^2=-A\left(x\right)dt^2+\frac{1}{B\left(x\right)}dx^2+x^2d\Omega^2\;.
\end{eqnarray}
Then Eqs.~(41) turn out to be
\begin{eqnarray}
&&K R_0^0-\frac{F}{2}+B K^{''}+\frac{B^{'}K^{'}}{2}+\frac{2BK^{'}}{x}=0\;,\\
&&K R_1^1-\frac{F}{2}+\frac{BA^{'}K^{'}}{2A}+\frac{2BK^{'}}{x}=0\;,\\
&&K R_2^2-\frac{F}{2}+B K^{''}+\frac{B^{'}K^{'}}{2}+\frac{BA^{'}K^{'}}{2A}\nonumber\\&&+\frac{BK^{'}}{x}=0\;,
\end{eqnarray}
where $K\equiv F_{,R}$ and prime denotes the derivative with respect to $x$. The components of Ricci tensor are given by
\begin{eqnarray}
&&R_0^0=\frac{-1}{4A^2x}\left(AA^{'}B^{'}x+2AA^{''}Bx-A^{'2}Bx+4AA^{'}B\right)\;,\nonumber\\
&&R_1^1=\frac{-1}{4A^2x}\left(AA^{'}B^{'}x+2AA^{''}Bx-A^{'2}Bx+4A^2B^{'}\right)\;,\nonumber\\
&&R_2^2=\frac{-1}{2Ax^2}\left(A^{'}Bx+AB^{'}x+2AB-2A\right)\;.
\end{eqnarray}
The combinations of Eqs.~(43-45) yield three equations of motion as follows
\begin{eqnarray}
&&K\left(R_0^0-R_1^{1}\right)+B K^{''}+\frac{B^{'}K^{'}}{2}-\frac{BA^{'}K^{'}}{2A}=0\;,\\
&&K\left(R_1^1-R_2^{2}\right)+\frac{BK^{'}}{x}-B K^{''}-\frac{B^{'}K^{'}}{2}=0\;,\\
&&K \left(R_2^2-R_0^{0}\right)+\frac{BA^{'}K^{'}}{2A}-\frac{BK^{'}}{x}=0\;.
\end{eqnarray}
Substituting Eqs.~(15-17) into Eqs.~(47-49), we find the solutions satisfy all the Einstein equations.
\section{understanding the solutions in the Einstein frame}
It is well-known that $F(R)$ gravity can be written in the Einstein frame where it emerges as General Relativity plus a scalar field. So in order to under the solutions better, let's analyze the solutions in the Einstein frame. The action for modified gravity in the absence of matter is given by

\begin{eqnarray}
S=\frac{1}{2\kappa^2}\int d^4x\sqrt{-g}F\left(R\right)\;,
\end{eqnarray}
where $\kappa^2=8\pi$. The action is equivalent to \cite{tey:83,wands:94}
\begin{eqnarray}
S=\frac{1}{2\kappa^2}\int d^4x\sqrt{-g}\left[F\left(\phi\right)+F^{'}\left(\phi\right)\left(R-\phi\right)\right]\;,
\end{eqnarray}
where $F^{'}\left(\phi\right)=dF/d\phi$. On can verify that the equation of motion for the scalar field $\phi$ gives
\begin{eqnarray}
\phi=R\;,
\end{eqnarray}
which reproduces the original action.   
Implemented with a conformal transformation,
\begin{eqnarray}
F^{'}\left(\phi\right)g_{\mu\nu}=\bar{g}_{\mu\nu}\;,
\end{eqnarray}
the action is reduced to the Einstein gravity plus the scalar field which is minimally coupled to gravity [32-35]
\begin{eqnarray}
S&=&\frac{1}{2\kappa^2}\int d^4x\sqrt{-\bar{g}}\left[\bar{R}-\frac{3}{2F^{'2}}\bar{g}^{\mu\nu}\bar{\nabla}_{\mu}F^{'}\bar{\nabla}_{\nu}F^{'}\right.\\&&\left.-\frac{1}{F^{'2}}\left(\phi F^{'}-F\right)\right]\;.
\end{eqnarray}
Define a canonical scalar field $\varphi$ by
 \begin{eqnarray}
\varphi=\sqrt{\frac{3}{2\kappa^2}}\ln F^{'}\;,
\end{eqnarray}
we find

\begin{eqnarray}
&&S=\int d^4x\sqrt{-\bar{g}}\left[\frac{1}{2\kappa^2}\bar{R}-\frac{1}{2}\bar{g}^{\mu\nu}\bar{\nabla}_{\mu}\varphi\bar{\nabla}_{\nu}\varphi-V\left(\varphi\right)\right]\;,\nonumber\\
&&V\left(\phi\left(\varphi\right)\right)=\frac{1}{2\kappa^2 F^{'2}}\left(\phi F^{'}-F\right)\;.
\end{eqnarray}
The corresponding Einstein equations are given by
\begin{eqnarray}
\bar{G}_{\mu\nu}=\kappa^2\left[-\bar{\partial}_{\mu}\varphi\bar{\partial}_{\nu}\varphi+\bar{g}_{\mu\nu}\left(\frac{1}{2}\bar{\nabla}^2\varphi-V\right)\right]\;.
\end{eqnarray}
In the next, we shall transform the solutions obtained in section II-C, from the Jordan frame to the Einstein frame. The first solution is for the Schwarzschild-de Sitter spacetime and it is trivial. So we shall start from the second one.

\subsubsection{$F=-2\sqrt{12c_2-R}$}
In this case, we derive the scalar field, the scalar field potential and the metric as follows  
\begin{eqnarray}
\varphi&=&\frac{\sqrt{6}}{4\kappa}\ln \frac{x^2}{2}\;,\nonumber\\
V&=&\frac{1}{2\kappa^2}e^{-\sqrt{6}\kappa\varphi}\left(1+12c_2e^{\sqrt{\frac{8}{3}}\kappa\varphi}\right)\;,\nonumber\\
ds^2&=&\bar{g}_{\mu\nu}dx^{\mu}dx^{\nu}\nonumber\\
&=&-\frac{\sqrt{2}}{2}x\left({1}+c_1x^{-2}+\frac{\sqrt{2}}{2}c_2x^2\right)dt^2+\frac{\sqrt{2}x^3}{4}d\Omega^2\nonumber\\&&+\frac{\sqrt{2}}{2}x\left({1}
    +c_1x^{-2}+c_2x^{2}\right)^{-1}dx^2\;.
\end{eqnarray}
We have checked they satisfy all the Einstein equations, Eq.~(58).

\subsubsection{$F={4}\cdot{5^{-\frac{5}{4}}}\cdot{R^{\frac{5}{4}}}$}
For the third solution presented in subsection $\textrm{C}$, we find the expressions of both the scalar potential and the Einstein tensor are rather lengthy.
So for simplicity, we shall take $c_2=0$ in this subsection. The corresponding solutions are
\begin{eqnarray}
\varphi&=&-\frac{\sqrt{6}}{8\kappa}\ln{2x^2}\;,\nonumber\\
V&=&\frac{1}{2\kappa^2}e^{\sqrt{6}\kappa\varphi}\;,\nonumber\\
ds^2&=&\bar{g}_{\mu\nu}dx^{\mu}dx^{\nu}\nonumber\\
&=&\frac{2^{-\frac{1}{4}}}{\sqrt{x}}\left[-x\left({1}
    +\frac{c_1}{x}\right)dt^2+\left({1}
    +\frac{c_1}{x}\right)^{-1}dx^2\right]\nonumber\\&&+\frac{2^{-\frac{1}{4}}}{\sqrt{x}}\cdot2x^2d\Omega^2\;.
\end{eqnarray}
We have checked they satisfy all the Einstein equations, Eq.~(58).

\subsubsection{$F=-\frac{81c_2^2}{R}-\frac{21}{2}$}
In this case, we have
\begin{eqnarray}
\varphi&=&\frac{\sqrt{6}}{\kappa}\ln{x}\;,\nonumber\\
V&=&\frac{3}{4\kappa^2}e^{-\sqrt{6}\kappa\varphi/2}\left(12c_2+7e^{-\kappa\varphi/\sqrt{6}}\right)\;,\nonumber\\
ds^2&=&\bar{g}_{\mu\nu}dx^{\mu}dx^{\nu}\nonumber\\
&=&-\frac{x^{\frac{7}{2}}+c_1+c_2x^{\frac{9}{2}}}{\sqrt{x}}dt^2+\frac{x^{\frac{11}{2}}}{x^{\frac{7}{2}}+c_1+c_2x^{9/2}}dx^2\nonumber\\&&+\frac{{4}}{7}x^4d\Omega^2\;,
\end{eqnarray}
and they satisfy all the Einstein equations, Eq.~(58).

\subsubsection{$F=\frac{2}{9}\left(-R\right)^{\frac{3}{2}}$}
In this case, we have
\begin{eqnarray}
\varphi&=&-\frac{\sqrt{6}}{4\kappa}\ln{\frac{x^2}{2}}\;,\nonumber\\
V&=&\frac{3}{2\kappa^2}e^{\sqrt{6}\kappa\varphi/3}\;,\nonumber\\
ds^2&=&\bar{g}_{\mu\nu}dx^{\mu}dx^{\nu}\nonumber\\
&=&\frac{\sqrt{2}}{{-x}}\left[-x^4\left({1}
    +\frac{c_1}{x}\right)dt^2+\left({1}
    +\frac{c_1}{x}\right)^{-1}dx^2\right]\nonumber\\&&+\frac{\sqrt{2}}{2}x d\Omega^2\;.
\end{eqnarray}
and find they satisfy all the Einstein equations, Eq.~(58).

\subsubsection{$F=-\frac{16}{R}-\frac{30}{49}c_2$}

In this case, we have
\begin{eqnarray}
\varphi&=&\frac{\sqrt{6}}{\kappa}\ln{\frac{x^2}{7}}\;,\nonumber\\
V&=&\frac{1}{49\kappa^2}e^{-\frac{2}{3}\sqrt{6}\kappa\varphi}\left(15c_2-196 e^{\kappa\varphi/\sqrt{6}}\right)\;,\nonumber\\
ds^2&=&\bar{g}_{\mu\nu}dx^{\mu}dx^{\nu}\nonumber\\
&=&-\frac{1}{49}x^8\left(-1+\frac{c_1}{x^7}+\frac{c_2}{x^2}\right)dt^2+\frac{{1}}{343}x^6d\Omega^2\nonumber\\&&+\frac{1}{49}\frac{x^4}{-1+\frac{c_1}{x^7}+\frac{c_2}{x^2}}dx^2\;.
\end{eqnarray}
and find they satisfy all the Einstein equations, Eq.~(58).  

Observing all the solutions presented above, we notice two remarkable properties in them. In the first place, the scalar field is not asymptotically vanishing. Secondly, the spacetime is not asymptotically flat. Hawking has argued that black holes cannot carry scalar charge, provided that the scalar field minimally couples to the metric \cite{haw:72}. That is to say scalar field has to be constant in the black hole spacetime. The conclusion has been generalized to the standard scalar-tensor
theories \cite{sot:12,bek:95,bek:96}. So the presence of the solutions obtained in this paper seems to violate the no-hair theorem. Then is there any inconsistency
with the theorem? The answer is no. We remember that one of the key assumptions in no-hair theorem is the spacetime of black holes should be asymptotically flat and the scalar field asymptotically vanishing. However, none of our solutions are asymptotically flat and the scalar field is asymptotically infinite. Actually, Sotiriou and Zhou \cite{sot:14} have pointed out that, in principle, one scalar field could always have a nontrivial configuration but without the black hole
carrying an extra (independent) scalar charge. This is sometime
referred to as ``hair of the second kind''.

\section{conclusion and discussion}\label{sec:conclusion}

In conclusion, we have developed a new method for looking for static and spherically symmetric solutions in $F(R)$ theory of gravity. The method consists of a right form of metric, Eq.~(5) and the right form of input functions, $f(r)$ or $K(r)$. With this method, the equations of motion become remarkably simple and solvable.

We have constructed a number of new solutions in terms of known analytic functions. It is found that there are always two event horizons and one curvature singularity in the spacetimes. Furthermore, a deficit angle always exists in the solutions except for the Schwarzschild-de Sitter solution. The other interesting points include the black hole spacetime is always asymptotically non-flat and the scalar field always asymptotically infinite. The two points violate the assumptions in the no-hair theorem. So the theorem doe not hold for these solutions. Finally, choosing different initial input functions, $f(r)$ or $K(r)$, one is expected to run into other new solutions.

\acknowledgments
We thank the two referees for the valuable suggestions. The work is supported by the Chinese MoST
863 program under grant 2012AA121701, the NSFC under grant
11373030, 10973014, 11373020 and 11465012.

\newcommand\ARNPS[3]{~Ann. Rev. Nucl. Part. Sci.{\bf ~#1}, #2~ (#3)}
\newcommand\AL[3]{~Astron. Lett.{\bf ~#1}, #2~ (#3)}
\newcommand\AP[3]{~Astropart. Phys.{\bf ~#1}, #2~ (#3)}
\newcommand\AJ[3]{~Astron. J.{\bf ~#1}, #2~(#3)}
\newcommand\APJ[3]{~Astrophys. J.{\bf ~#1}, #2~ (#3)}
\newcommand\APJL[3]{~Astrophys. J. Lett. {\bf ~#1}, L#2~(#3)}
\newcommand\APJS[3]{~Astrophys. J. Suppl. Ser.{\bf ~#1}, #2~(#3)}
\newcommand\JHEP[3]{~JHEP.{\bf ~#1}, #2~(#3)}
\newcommand\JMP[3]{~J. Math. Phys. {\bf ~#1}, #2~(#3)}
\newcommand\JCAP[3]{~JCAP. {\bf ~#1}, #2~ (#3)}
\newcommand\LRR[3]{~Living Rev. Relativity. {\bf ~#1}, #2~ (#3)}
\newcommand\MNRAS[3]{~Mon. Not. R. Astron. Soc.{\bf ~#1}, #2~(#3)}
\newcommand\MNRASL[3]{~Mon. Not. R. Astron. Soc.{\bf ~#1}, L#2~(#3)}
\newcommand\NPB[3]{~Nucl. Phys. B{\bf ~#1}, #2~(#3)}
\newcommand\CMP[3]{~Comm. Math. Phys.{\bf ~#1}, #2~(#3)}
\newcommand\CQG[3]{~Class. Quant. Grav.{\bf ~#1}, #2~(#3)}
\newcommand\PLB[3]{~Phys. Lett. B{\bf ~#1}, #2~(#3)}
\newcommand\PRL[3]{~Phys. Rev. Lett.{\bf ~#1}, #2~(#3)}
\newcommand\PR[3]{~Phys. Rep.{\bf ~#1}, #2~(#3)}
\newcommand\PRD[3]{~Phys. Rev. D{\bf ~#1}, #2~(#3)}
\newcommand\RMP[3]{~Rev. Mod. Phys.{\bf ~#1}, #2~(#3)}
\newcommand\SJNP[3]{~Sov. J. Nucl. Phys.{\bf ~#1}, #2~(#3)}
\newcommand\ZPC[3]{~Z. Phys. C{\bf ~#1}, #2~(#3)}
 \newcommand\IJGMP[3]{~Int. J. Geom. Meth. Mod. Phys.{\bf ~#1}, #2~(#3)}
  \newcommand\GRG[3]{~Gen. Rel. Grav.{\bf ~#1}, #2~(#3)}
  \newcommand\EPJC[3]{~Eur. Phys. J. C{\bf ~#1}, #2~(#3)}

\end{document}